\documentclass{appolb}
\usepackage{epsfig}

\begin{document}

\title{GENIUS project, neutrino oscillations and Cosmology: 
neutrinos reveal their nature?\thanks{Presented by M. Zra\l ek 
at the Cracow Epiphany Conference on Neutrinos in Physics and Astrophysics, 
January 6-9, 2000, Cracow, Poland.
}$\;$\thanks{Work supported in part by the
Polish Committee for Scientific Research under 
Grant No.~~2P03B05418. 
J.G. would like to thank the Alexander von Humboldt-Stiftung for 
fellowship.
}}
\author{M. CZAKON, J. STUDNIK AND M. ZRA\L EK
\address {Department of Field Theory and Particle Physics, Institute 
of Physics, \\ 
University of
Silesia, Uniwersytecka 4, PL-40-007 Katowice, Poland} \\ \vspace{.5cm}
J. GLUZA
\address {Department of Field Theory and Particle Physics, Institute 
of Physics, \\ University of
Silesia, Uniwersytecka 4, PL-40-007 Katowice, Poland, \\
DESY Zeuthen, Platanenallee 6, 15738 Zeuthen, Germany}
}
\maketitle

\begin{abstract}
 
The neutrinoless double beta decay as well as any other laboratory experiment has not been able to answer the question of the 
neutrino's nature.
Hints on the answer are available when neutrino oscillations and $( \beta \beta  )_
{0 \nu}$ are considered simultaneously. In this case phenomenologically  interesting
neutrino mass schemes can lead to non-vanishing and large values of $\langle m_{\nu} \rangle$. As a consequence,
some schemes with Majorana neutrinos can be ruled out even now.
If we assume that in addition neutrinos contribute to Hot Dark Matter then the window for Majorana neutrinos is even
more restricted, e.g. GENIUS experiment will be sensitive to  scenarios with three Majorana neutrinos.

\end{abstract}
\section{Introduction}

There are two main problems in neutrino physics. First is the problem of
neutrino mass. In the light of present observations \cite{Ref1} this question seems
to be solved, neutrinos are massive particles. The second problem is the one of the neutrino's nature.
Massive neutrinos can be either Dirac or Majorana particles. As their
visible interactions are left-handed and known sources generate
ultrarelativistic states, it is very difficult to distinguish 
experimentally between the two.

Dirac spin 1/2 fermions were introduced  to describe interactions which
are invariant under spatial reflections. Majorana fermions were invented
later  without special purpose. In those times, parity was conserved and it
was generally believed that Majorana particle interactions must be asymmetric
under spatial reflection. Today we know that particle characters are not
responsible for parity symmetry breaking \cite{Ref4}. Only one property - the charge
- discriminates Dirac from Majorana massive fermions. Dirac particles carry
charge. Massive Majorana particles must be chargeless and cannot carry 
static electric or magnetic  moments.

Neutrinos are ``special'' fermions, they have no electric charge and only one
``charge'' - the lepton number can characterize them. From all present
terrestrial experiments it follows that family lepton numbers $L_e$, $L_\mu $
and $L_\tau $ are separately conserved and as a result, their sum, the total
lepton number $L=$ $L_e$ $+L_\mu $ $+$ $L_\tau $ has the same property.

For massive Dirac neutrinos, flavor lepton numbers can be broken and only L
must be conserved. For Majorana neutrinos both, family and total lepton
numbers are broken. It is even impossible to define these numbers in the way
known from Dirac particles.

We have to stress that Majorana neutrinos are more fundamental objects and
naturally arise in most extensions of the Standard Model. Only in models
where the lepton number ($L$ and $B-L$) is conserved,
neutrinos are Dirac particles. However, there are many arguments to
abandon lepton number conservation. It is not a fundamental quantity,
unlike electric charge and does not govern the dynamics. 
Also, lepton number violation is naturally
induced by the presence of right-handed neutrinos $\left( \nu _R\right) $
which are usually necessary to form the Dirac mass term $\left( \overline{\nu }%
_L\nu _R\right)$. In spite of these obvious theoretical arguments,
supporting the Majorana nature of neutrinos, finding  direct experimental
indications which would determine the neutrino character is very
important.

It is common belief that the first place where the nature of massive
neutrinos will be revealed is the neutrinoless double $\beta $ decay of nuclei, $%
\left( \beta \beta \right) _{0\nu }$. Many experimental searches for $%
\left( \beta \beta \right) _{0\nu }$ decay of different nuclei have been done and are
presently underway \cite{Ref5}. Unfortunately, up to now this decay has not been found and the
experimental data can only help to estimate the lower bound on the life times of $\left(
\beta \beta \right) _{0\nu }$ decay modes. The most
stringent limit was found in the germanium Heidelberg-Moscow experiments.
Their latest result on the half-life time is \cite{Ref6} 
\begin{equation}
T_{1/2}^{o\nu }\left( Ge\right) >5.7\times 10^{25}\mbox{ year (at 90\% CL)}
\end{equation}
from which the following upper bound on the effective Majorana mass was found 
\begin{equation}
|\left\langle m_\nu\right\rangle |=\left| \sum_iU_{ei}^2m_i\right| <0.2 \;\; eV.
\end{equation}
The above number has been used to restrict many aspects of the neutrino mass spectrum
or the solar neutrino mechanism \cite{Ref7}. We propose an opposite way of thinking.
Using the present data from oscillation experiments, and tritium $\beta $
decay we can find the modules $\left| U_{ei}\right| $ of mixing matrix
elements and the possible values of neutrino masses $m_i$. Then we check whether
the bound Eq. (2) is satisfied or not. If it is, the problem is
unsolved. If, however
the bound Eq. (2) is not satisfied, then neutrinos
are Dirac particles. In the latter case the effective mass is calculated as \cite{Ref8} 
\begin{equation}
\sum_{l=1}^n U_{el}^2m_l\Rightarrow \sum_{l=1}^n\frac 12\left[ \left(
-iU_{el}\right) ^2+\left( U_{el}\right) ^2\right] m_l=0
\end{equation}
and is strictly equal zero.

Ambitious plans \cite{gen} are to shift up the limit Eq. (1) and to move the upper limit of $<m_{\nu}>$
down to 0.02 eV (using a tank of 1 ton of Germanium, after one year) or in a further time scale 
even to 0.006 eV (1t, 10 years).

In a previous work  \cite{Ref9} we have considered three neutrino mixing schemes.
Here we present analytical results for both three and four neutrino mixing scenarios.
Information about the same subject with numerical estimations is given in \cite{Ref10}.
In the next Chapter we summarize the efforts undertaken in order to find lepton number
violating processes. Explanation is given of why the family and total lepton
numbers are so strongly conserved. In Chapter 3 we collect all the relevant
information about mixing matrix elements and masses extracted from
experimental data. Four presently accepted neutrino mass schemes which cover the
case of three and four neutrino mixing are discussed. All necessary
information from oscillation experiments, tritium $\beta $ decay and
cosmology is given. Chapter 4 is the main of the paper. All data are
connected together with the bound on effective neutrino mass from 
$(\beta \beta) _{0\nu }$, and restrictions on various neutrino mass
schemes are presented. Conclusions are to be found in Chapter 5.

\section{Lepton numbers and neutrino character of light SM neutrino states}

In order to explain the lack of lepton flavor violating processes, the concept of the 
flavor lepton number $L_{\alpha}$ \cite{Ref11} followed by the idea of the
total lepton number $L$ \cite{Ref12}  have been introduced.
The upper bounds on branching ratios of  $L_{\alpha}$ violating processes are very small, for instance:
\begin{equation}
\begin{array}{ll}
BR( \mu^- \to e^- \gamma ) < 4.9 \cdot 10^{-11}, & BR( \mu^- \to e^+e^-e^- ) < 1.0 \cdot 10^{-12}, \cr
& \cr
BR( \pi^0 \to \mu^- e^+) < 1.72 \cdot 10^{-8}, &
BR( K_L^0 \to \mu^- e^+ ) < 3.3 \cdot 10^{-11}, \cr
& \cr
BR( \tau^- \to \mu^- \gamma) < 4.2 \cdot 10^{-6}. &
\end{array}
\label{br}
\end{equation}
In the frame of the SM with  massless neutrinos the above processes are strictly forbidden.
If neutrinos are massive,  then  in analogy to the quark sector
neutrinos should mix and lepton numbers are not conserved. However, these effects  must be very small,
below sensitivity of processes given in Eq.~(\ref{br}). That means that the concept of 
leptons numbers  $L,L_{\alpha}$ is  still usefull, at least  in all neutrino nonoscillation phenomena.
For Dirac neutrinos represented by a bispinor $\Psi_D$
it is possible to change the phase of the field
\begin{equation}
  \Psi_D \to e^{i \alpha} \Psi_D.
\label{dir}
\end{equation}
The charge connected with such a global gauge transformation is just the flavor charge operator. 
This operator can, but not necessarily must, commute with the interaction Hamiltonian, $[L_{\alpha},H]=0$
for a massless, $[L_{\alpha},H] \neq 0$ for a massive neutrino.
Majorana neutrinos on the other hand are described by self-conjugate fields
 \begin{equation}
  \Psi_M =\Psi_M^c \equiv C \bar{\Psi}^T_M,
\end{equation}
and it is not possible to define the same kind of gauge transformation as in Eq.~(\ref{dir}). 
There is then no special reason why $L_{\alpha}$ and $L$ should be conserved for Majorana neutrinos.
All processes in Eq.~(\ref{br}) break  $L_{\alpha}$ but not $L$, so they can be 
realized by both kind of neutrinos  at
the one loop level. At this level only very heavy,  nonstandard, neutrinos matter \cite{lh}.
We do not go to  details and focus only on direct effects connected  with light, SM neutrinos. 
Let us mention only that in see-saw models heavy neutrino effects are also negligible, both at tree 
\cite{tr} and loop levels \cite{ll}. 
To make life easier and to  understand  how processes with Majorana (Dirac)
neutrinos mimic
family $L_{\alpha}$ and total lepton $L$ numbers conservation let us consider 
a tree level process  of electron (positron) production using electron and muon neutrinos
scattering on nuclear target
\begin{equation}
\nu_{e(\mu)} N \to e^{\pm } X.
\label{proc}
\end{equation}

Let us define the connection between flavor $\nu_{\alpha}$ and massive $\nu_i$ states in the following way
\begin{equation}
|\nu_{\alpha} (\lambda=-\frac{1}{2})\rangle=\sum\limits_i U_{\alpha i} |\nu_{i} (\lambda=-\frac{1}{2})\rangle
\label{par}
\end{equation}
for negative helicity states and 
\begin{equation}
|\nu_{\alpha} (\lambda=+\frac{1}{2})\rangle=\sum\limits_i U_{\alpha i}^{\ast} |\nu_{i} (\lambda=+\frac{1}{2})\rangle
\label{antpar}
\end{equation}
for $\lambda=+\frac{1}{2}$. In the same way Weyl particle and antiparticle states are connected.
Note that Eqs.~(\ref{par}-\ref{antpar}) for massive particles seems to be in contradiction 
with  the special theory of relativity. 
The real problem is that states on the right hand side of  Eqs.~(\ref{par}-\ref{antpar}) can not be defined, in general
 \cite{fock}. However,
the left-handed interaction cannot change the neutrino helicity and it is practically impossible 
to find a real frame moving
faster than the neutrino itself (neutrinos are ultrarelativistic) and  the relations  
Eqs.~(\ref{par}-\ref{antpar}) can be safely used \cite{fock}. 
This is what is usually considered to be true when neutrino oscillation phenomena are discussed.
To be more general, let us assume  that there is also a right-handed neutrino interaction
\footnote{For Dirac neutrinos there are actually two independent neutrino mixing matrices in left- and right-
handed charged currents \cite{ac}. Even for Majorana neutrinos these could be in principle different (as 
is e.g. the case of see-saw type models where the light neutrino mixing matrix in the right-handed current is 
dumped by the heavy neutrino mass scale). These simplifications do not spoil the general idea given here.}
\begin{eqnarray}
L_{CC}&=&\frac g{\sqrt{2}}\left[ A_L\left( \overline{N}_i \gamma ^\mu
P_L (U^T)_{i \alpha} l_{\alpha} \right) W_{L \mu}^+ +A_R\left( \overline{N}_i \gamma ^\mu P_R
{(U^{\dagger})}_{i \alpha} l_{\alpha} \right) \right] W_{R \mu}^{+}
+h.c. \nonumber \\
&&
\label{cc}
\end{eqnarray}
Then in the ultrarelativistic regime  $(m_i << E)$ using the unitarity of the U matrix, 
the following  amplitudes  to the 
${\cal{O}} \left( \left( \frac{m_i}{E} \right)^2 \right)$ order are obtained:
\begin{eqnarray}
A \left( \nu_e (-1/2) \to e^- \right) &=& A(e^-) \left[ A_L^{\ast}+A_R^{\ast} \sum\limits_i \frac{m_i}{2E}
                     \left( U_{ei} \right)^2 \right], \label{min} \\
A \left( \nu_e (+1/2) \to e^- \right) &=& A(e^-) \left[ A_L^{\ast} \sum\limits_i \frac{m_i}{2E}
                     \left( U_{ei}^\ast \right)^2 +A_R^{\ast}\right], \label{t1} \\
A \left( \nu_\mu (-1/2) \to e^- \right) &=& A(e^-) \left[ -A_L^{\ast}\sum\limits_i \frac{m_i^2}{8E^2}
                       U_{\mu i} U_{ei}^{\ast}+A_R^{\ast} \sum\limits_i \frac{m_i}{2E} 
                       U_{\mu i} U_{ei} \right], \nonumber \\ 
&& \label{3a} \\
A \left( \nu_\mu (+1/2) \to e^- \right) &=& A(e^-) \left[ A_L^{\ast}\sum\limits_i \frac{m_i}{2E}
                    U_{\mu i}^{\ast} U_{ei}^{\ast}-A_R^{\ast} \sum\limits_i \frac{m_i^2}{8E^2}
                      U_{\mu i}^{\ast} U_{ei}  \right] \label{4a} 
\end{eqnarray}
and
\begin{eqnarray}
A \left( \nu_e (-1/2) \to e^+ \right) &=& A(e^+) \left[ -A_L  \sum\limits_i \frac{m_i}{2E} 
                       \left( U_{ei} \right)^2 +A_R  \right], \label{t2} \\
A \left( \nu_e (+1/2) \to e^+ \right) &=& A(e^+) \left[- A_L 
                      +A_R \sum\limits_i \frac{m_i}{2E}\left( U_{ei}^\ast \right)^2 \right],\label{pl} \\
A \left( \nu_\mu (-1/2) \to e^+ \right) &=& A(e^+) \left[ -A_L \sum\limits_i \frac{m_i}{2E}
                    U_{\mu i} U_{ei}+A_R \sum\limits_i \frac{m_i^2}{8E^2}
                      U_{\mu i} U_{ei}^{\ast}  \right] , \nonumber \\
&& \label{3b} \\ 
A \left( \nu_\mu (+1/2) \to e^+ \right) &=& A(e^+) \left[ A_L\sum\limits_i \frac{m_i^2}{8E^2}
                       U_{\mu i}^{\ast} U_{ei}+A_R^{\ast} \sum\limits_i \frac{m_i}{2E} 
                       U_{\mu i}^{\ast} U_{ei}^{\ast} \right] \label{4b}
\end{eqnarray}
where $ A(e^+)$ and $ A(e^-)$ are appropriate amplitudes for massless neutrinos. In the approximation 
$\frac{m_i}{E}<<1$ and $|A_L|>>|A_R|$ only two cross sections for electron production by a 
$\nu_e (\lambda=-1/2)$ beam Eq.~(\ref{min}) and positron production by a 
 $\nu_e (\lambda=+1/2)$ beam Eq.~(\ref{pl}) are large enough to be seen 

\begin{equation}
\sigma \left( \nu_e (-1/2) \to e^- \right) \sim  |A(e^-)|^2,
\end{equation}
\begin{equation}
\sigma \left( \nu_e (+1/2) \to e^+ \right) \sim  |A(e^+)|^2.
\end{equation}
All other helicity cross sections are suppressed by factors
\begin{equation}
\left(  \frac{m_i}{E} \right)^2, \;\;\;  \frac{m_i}{2E}|A_R| \;\;\; \mbox{\rm or} \;\;  |A_R|^2,
\label{fac}
 \end{equation}
and for instance, for $m_i \simeq 1$ eV and $E \simeq 1$ MeV we have  
$\left(  \frac{m_i}{E} \right)^2 \simeq 10^{-12}$.
Such factors cause that the cross sections for flavor 
lepton number $L_{\alpha}$  violating processes  Eqs.~(\ref{3a}-\ref{4a}),  Eqs.~(\ref{3b}-\ref{4b}) 
are invisibly small. 
The total lepton $L$
non-conserving processes Eqs.~(\ref{t1},\ref{t2}), share the same property.
Neglecting the factors from Eq.~(\ref{fac}), our amplitudes are identical to those of massless Weyl neutrinos whose
family and total lepton numbers are strictly conserved.
Turning our results around we can see 
that processes where neutrino masses (and right-handed currents) are not important
give no chance to distinguish Dirac from Majorana neutrinos. Could CP phases help?
In the case of Dirac [Majorana] neutrinos the mixing matrix U has $(n-1)(n-2)/2 \left[ n(n-1)/2 \right]$
phases. Let us look into processes where the neutrino mass is important.  
Though  the transition probability  of neutrino oscillations depends on CP phases, 
the physical phases by which the neutrino mixing matrices differ do not enter into transition probabilities
and  the results are the same  for Dirac and Majorana 
neutrinos \cite{ac,a2}. The neutrino mass distortion measured in tritium $\beta$ decay is a function of absolute values 
of mixing matrix elements (see next chapter)
so it is not sensible to CP phases, either.

There are also processes which do not conserve the total lepton number in which only Majorana neutrinos could
participate.
Since many years the most promising
investigation along this line is connected with the neutrinoless double beta decay. 

Surprisingly, we will see that even if this process is not observed, it can solve the problem of the
nature of neutrinos, when augmented with Cosmology (assuming  neutrinos as Hot Dark Matter) 
and neutrino oscillations results.

\section{Neutrino masses and mixing matrix $U_{ei}$ elements}

There are two completely different situations which depend on the present status of the LSND experiment. Three
light neutrinos are necessary to explain solar \cite{Ref13} and atmospheric \cite{Ref14} anomalies. With the LSND
result \cite{Ref15} an extra light neutrino must be introduced. 

\subsection{Three neutrinos scenario}

For neutrino mixing  3 flavor states $(\nu
_e,\nu _\mu ,\nu _\tau )$ are related to 3 eigenmass states $(\nu _1,\nu _2,\nu _3)$ through\cite{Ref16}%
\begin{equation}
\left( 
\begin{array}{c}
\nu _e \\ 
\nu _\mu \\ 
\nu _\tau 
\end{array}
\right) =\left( 
\begin{array}{ccc}
U_{e1} & U_{e2} & U_{e3} \\ 
U_{\mu 1} & U_{\mu 2} & U_{\mu 3} \\ 
U_{\tau 1} & U_{\tau 2} & U_{\tau 3} 
\end{array}
\right) \left( 
\begin{array}{c}
\nu _1 \\ 
\nu _2 \\ 
\nu _3 
\end{array}
\right) . 
\end{equation}
Our concern is about the first row of the mixing matrix. We use the standard parameterization \cite{Ref17}
\begin{equation}
U=\left( 
\begin{array}{ccc}
c_{12}c_{13} & s_{12}c_{13} & s_{13}e^{i \delta} \\ 
-s_{12}c_{23}-c_{12}s_{23}s_{13}e^{-i \delta} & c_{12}c_{23}-s_{12}s_{23}s_{13}e^{-i\delta} & s_{23}c_{13} \\ 
s_{12}s_{23}-c_{12}c_{23}s_{13}e^{-i \delta} & -c_{12}s_{23}-s_{12}c_{23}s_{13}e^{-i\delta} & c_{23}c_{13} 
\end{array}
\right) 
\end{equation}
To explain the solar neutrino anomaly the  mass splitting between two neutrinos must be extremely tiny,
$\delta m_{sun}^2 \simeq 10^{-5} \div 10^{-11}\; eV^2$. Only slightly larger mass splitting between neutrino masses
is needed in the case of atmospheric oscillations, $\delta m_{atm}^2 \simeq 10^{-2} \div 10^{-3}\; eV^2$.
These relations leave us with two possible neutrino mass scenarios:
$\delta m_{12}^2 =\delta m_{sun}^2$ ,    $\delta m_{23}^2 \simeq \delta m_{13}^2 = \delta m_{atm}^2$ (Scheme $A_3$,
Fig.1) and  $\delta m_{23}^2 =\delta m_{sun}^2$,  $\delta m_{12}^2 \simeq \delta m_{13}^2 = \delta m_{atm}^2$ 
(Scheme $B_3$, Fig.1) 
\begin{figure}
\epsfig{figure=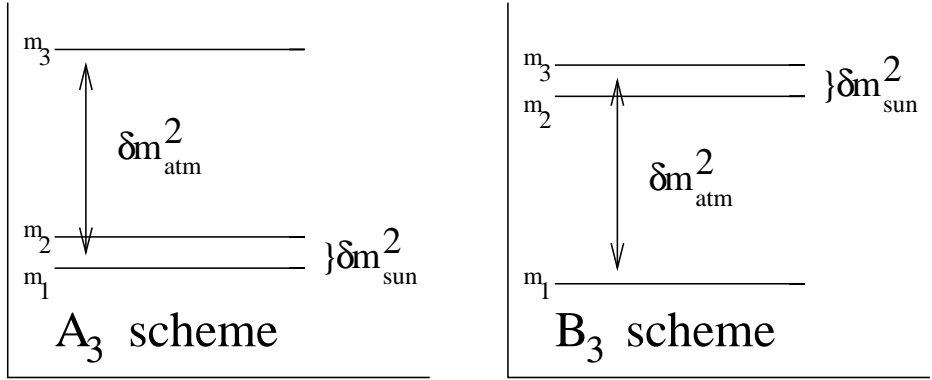, height=2 in}
\caption{Two possible neutrino mass spectra which can describe the 
oscillation data.}
\end{figure}
Reactor experiments are of the so-called short baseline and are able to measure the neutrino mass splitting of the order 
$\delta m^2 < 10^{-3}$. Then we can neglect terms with $\delta m_{sun}^2 \ll 10^{-3}$ and the disappearance 
probability for $\bar{\nu}_e$ reactor neutrino oscillations is (the following discussion is given for the
$A_3$ scheme)
\begin{equation}
P_{\bar{\nu}_e \to \bar{\nu}_e}=P_{{\nu}_e \to {\nu}_e}=1-\sin^2{2 \Theta_{13}} \sin^2{\Delta_{reactor}},
\end{equation}
where
$$
\Delta_{reactor}=
\Delta_{23} (L_{reactor},E_{reactor}), \;\;\; \Delta_{ij} (L,E)=\frac{1.27 \times \delta m^2_{ij} [eV^2] L [km]}{E [GeV]} .
$$
In reactor experiments the disappearance of $\bar{\nu}_e$ is not seen, which means that $\sin^22\Theta_{13}$ must be
small. CHOOZ gives \cite{chooz}
\begin{equation}
\sin^2{2 \Theta_{13}} < 0.18,
\end{equation}
and two solutions for $\Theta_{13}$ can be found:
\begin{equation}
\sin^2{ \Theta_{13}} <0.05 \;\;\; \mbox{\rm or} \;\;\; \sin^2{ \Theta_{13}}> 0.95.
\label{small}
\end{equation}
The observed ${\nu}_{\mu}$ neutrino deficit from the atmosphere is favorable describe by a
${\nu}_{\mu} \to {\nu}_{\tau}$ transition where matter effects are not important
$\left( \Delta_{atm}= \Delta_{23} (L_{atm},E_{atm}) \right)$: 
\begin{equation}
P_{\nu_{\mu} \to {\nu}_{\tau}}=\sin^2{2 \Theta_{23}} \cos^4{ \Theta_{13}} \sin^2{ \Delta_{atm}}.
\end{equation}
We know that the atmospheric neutrino mixing is very large \cite{Ref18}
\begin{equation}
0.72 \leq \sin^2{2 \Theta_{23}} \cos^4{ \Theta_{13}}  \leq 1 \;\;\; \mbox{\rm and} \;\;\; \delta m_{atm}^2 \simeq
4 \times 10^{-3} \; eV^2
\label{con}
\end{equation}
and only a small value of $\sin^2{ \Theta_{13}}$ in Eq.~(\ref{small}) is compatible with the bound in Eq.~(\ref{con}).
 The recent fit to the new (830-920 days) atmospheric data of Superkamiokande gives the minimum of  $\chi^2$ for
\cite{Ref28}
\begin{equation}
\sin^2{ \Theta_{13}} =0.03. 
\label{best}
\end{equation}
Similar values ($\sin^2{ \Theta_{13}}<0.03 \div 0.04$ \cite{chooz,ref25}) 
are given by the reactor data.

In all solar neutrino experiments the deficit of electron antineutrinos is measured and four different solutions
are possible \cite{Ref19}. The first, ``just so'' solution, is based on the hypothesis of neutrino 
oscillations in vacuum (VO), 
 $\delta m_{sun}^2 \sim 10^{-10}\;eV^2$ in this case. The other three solutions are based on the Wolfenstein\cite{Ref20}-Mikheyev-Smirnov
\cite{Ref21} mechanism of coherent neutrino scattering in matter (so called small mixing angle (SMA MSW), 
large mixing angle (LMA MSW) and low $\delta m^2$ (LOW MSW)  solutions). 

For VO the $\nu_e$ disappearance probability is given by $\left( {\Delta_{sun}}=
\Delta_{12} (L_{sun},E_{sun}) \right)$   
\begin{eqnarray}
P_{{\nu}_e \to {\nu}_e}^{sun}&=&1-\frac{1}{2} \sin^2{2 \Theta_{13}} -\sin^2{2 \Theta_{12}}
\cos^4{ \Theta_{13}} \sin^2{\Delta_{sun}}.
\end{eqnarray}
This expression can be rewritten in the form 
\begin{equation}
P_{{\nu}_e \to {\nu}_e}^{sun}=\cos^4{ \Theta_{13}}  \left( 1-\sin^2{2 \Theta_{12}} \sin^2{\Delta_{sun}} \right)
+ \sin^4{ \Theta_{13}}.
\label{3}
\end{equation}
Taking into account that $ \sin^4{ \Theta_{13}} \simeq 0$ (Eq.~(\ref{best})) we get
\begin{equation}
P_{{\nu}_e \to {\nu}_e}^{sun} \simeq 1-\sin^2{2 \Theta_{sun}}  \sin^2{\Delta_{sun}}.
\end{equation}
where $ \Theta_{sun} \simeq  \Theta_{12}$.

Similarly we get for the case of the MSW solution:
\begin{equation}
P_{{\nu}_e \to {\nu}_e}^{sun(MSW)} \simeq 1-\sin^2{2 \tilde{\Theta}_{sun}}  \sin^2{\tilde{\Delta}_{sun}}.
\end{equation}
where now
\begin{equation}
\sin^2{2 \tilde{\Theta}_{sun}}=\frac{\sin^2{2 {\Theta}_{sun}}} {\left[ \left( \frac{A}{\delta m^2_{sun}}-
\cos{2 \Theta_{sun}} \right)^2+\sin^2{2 {\Theta}_{sun}} \right]^{1/2}}.
\end{equation}
$\tilde{\Delta}_{sun}$ includes the effective neutrino mass parameter with $\delta m^2_{sun}$ replaced
by
\begin{equation}
\tilde{\delta} m^2_{sun}= \delta m^2_{sun} \left[ \left(  \frac{A}{\delta m^2_{sun}}-
\cos{2 \Theta_{sun}} \right)^2 + \sin^2{2 {\Theta}_{sun}} \right]^{1/2},
\label{res}
\end{equation}
where $A=2 \sqrt{2} G_F E N_e$ ($N_e$ - electron number density).

From Eq.~(\ref{res}) we can see that in order to fulfill the resonance condition we need for $\delta m^2_{sun} > 0$  \cite{Ref32}
\begin{equation}
\cos{2 \Theta_{sun}} > 0,
\end{equation}
and this means that 
\begin{equation}
\cos{ \Theta_{12}} > \sin{ \Theta_{12}}.
\label{st}
\end{equation}

Many fits have been done to the solar neutrino data \cite{wszystkie_fity}. The results of the fit \cite{Ref26} 
which takes into account the full 
set of measurements (rates, energy spectrum, day-night asymmetry in the case of the MSW solution and seasonal 
variation for VO solution) are presented in 
Table I. For VO only  the best fit value  $\sin^2{2 \Theta_{sun}}$ is given in \cite{Ref27}.
\begin{table}
\begin{tabular}{|c|c|c|c|}\hline
\label{mnueff}
Possible solutions  & $\sin^2{2 \Theta_{sun}}$ [95 \% c.l.]& \multicolumn{2}{|c|}{Best fits}  \\ 
\cline{3-4}  
& & 
$\sin^2{2 \Theta_{sun}}$  &  $\delta m^2$ \\
\hline
MSW SMA & 0.001 - 0.01 & 0.0065 & $5.2 \times 10^{-6} eV^2$ \\
\hline
MSW LMA & 0.59 - 0.98 & 0.77 & $2.94 \times 10^{-5} eV^2$ \\
\hline
MSW LOW & 0.68-0.98 & 0.9 & $1.24 \times 10^{-7} eV^2$ \\
\hline
VO &  & 0.93 & $4.42 \times 10^{-10} eV^2$ \\
\hline
\end{tabular}
\caption{The allowed ranges and best fit values of $\sin^2{2 \Theta_{sun}}$ and $\delta m^2$ for different 
types of solar neutrino oscillations.}
\end{table}

For a scheme $B_3$ a change $U_{e3} \leftrightarrow U_{e1}$ must be done.
\subsection{Four neutrinos scenario}

The electron (anti)neutrino appearance in the LSND experiment \cite{Ref15,Ref31} can be explained by 
$\nu_{\mu} \to \nu_e$ oscillation with additional large  $\delta m^2$ scale
\begin{equation}
 \delta m^2_{LSND} \sim 0.2 \div 2\; \mbox{\rm eV}^2.
\end{equation}
In principle there are six possible four-neutrino mass schemes with three different scales of $\delta m^2$.
They are widely discussed in literature \cite{schw} and it is known that only two schemes (Fig.2) are accepted
by reactor, LSND, solar and atmospheric neutrino data.

\begin{figure}
\epsfig{figure=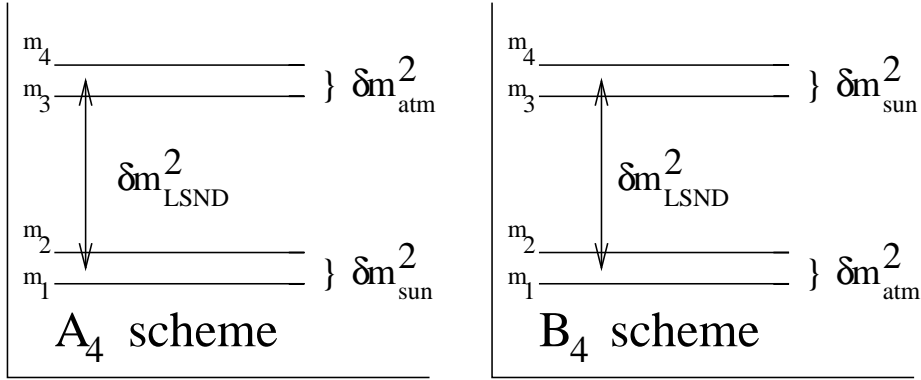, height=2 in}

\caption{Two accepted by present data four-neutrino schemes. In the scheme $A_4$ $\delta m^2_{sun}=
\delta m^2_{12}$,  $\delta m^2_{atm}=\delta m^2_{34}$ with opposite situation in $B_4$ scheme where 
$\delta m^2_{sun}=\delta m^2_{34}$,  $\delta m^2_{atm}=\delta m^2_{12}$. In both schemes 
$\delta m^2_{LSND} \simeq \delta m^2_{13} \simeq \delta m^2_{14}  \simeq \delta m^2_{23}
 \simeq \delta m^2_{24}$.}
\end{figure}

As the parameterization of the $4 \times 4$ neutrino mixing matrix is very complicated \cite{Ref33} in the case when 
all entries of the mass matrix are nonzero  we will use only the symbolic denotations and take

\begin{equation}
\left( 
\begin{array}{c}
\nu _e \\ 
\nu _\mu \\ 
\nu _\tau \\
\nu_s 
\end{array}
\right) =\left( 
\begin{array}{cccc}
U_{e1} & U_{e2} & U_{e3} &  U_{e4} \\ 
U_{\mu 1} & U_{\mu 2} & U_{\mu 3}& U_{\mu 4} \\ 
U_{\tau 1} & U_{\tau 2} & U_{\tau 3} & U_{\tau 4} \\
U_{s1} & U_{s2} & U_{s3} &  U_{s4}
\end{array}
\right) \left( 
\begin{array}{c}
\nu _1 \\ 
\nu _2 \\ 
\nu _3  \\
\nu_4
\end{array}
\right) . 
\end{equation}
For the short baseline experiment (and for scheme $A_4$) the  probability of disappearance of $\nu_e$ neutrinos
is given by $\left(  \Delta_{SBL}=\Delta_{32}(L_{SBL},E_{SBL}) \right)$
\begin{equation}
P_{{\nu}_e \to {\nu}_e}=1-c_e (1-c_e) \sin^2{\Delta_{SBL}}
\end{equation}
where
$$c_e=|U_{e3}|^2+|U_{e4}|^2.
$$
Again we can implement the CHOOZ result \cite{chooz} and get

\begin{equation}
4 c_e (1-c_e) < 0.18,
\end{equation}
and there are two solutions for $c_e$, namely
\begin{equation}
 c_e < 0.05 \;\;  \mbox{\rm or} \;\;  c_e > 0.95.
\label{one}
\end{equation}
On the other hand the deficit of solar neutrinos in the VO scenario and in four-neutrino language reads 
$\left( \Delta_{sun}=\Delta_{12}(L_{sun},E_{sun}) \right)$ 
\begin{equation}
P_{{\nu}_e \to {\nu}_e}^{(4)}= \left( 1-|U_{e3}|^2-|U_{e4}|^2 \right)^2 \left[ 
1-\frac{4|U_{e1}|^2|U_{e2}|^2}{\left( |U_{e1}|^2+|U_{e2}|^2 \right)^2} \sin^2{\Delta_{sun}} \right]
+|U_{e3}|^4+|U_{e4}|^4.
\end{equation}

By comparison with the 
 two flavour oscillations formula
\begin{equation}
P_{{\nu}_e \to {\nu}_e}=1-\sin^2{2 \Theta_{sun}}\sin^2{\Delta_{sun}}
\end{equation}
we can see that the factor $\left( 1-|U_{e3}|^2-|U_{e4}|^2 \right)^2$ must be close to one, so only one solution of 
Eq.~(\ref{one}) is possible, namely $c_e<0.05$. For small values of $c_e$ the probability 
$P_{{\nu}_e \to {\nu}_e}^{(4)}$ is well described in the frame
of the 3 neutrino scenario by $P_{{\nu}_e \to {\nu}_e}^{(3)}$ (Eq.~(\ref{3})) with the following substitutions
\begin{eqnarray}
|U_{e3}|^2+|U_{e4}|^2 &  \to & \sin^2{\Theta_{13}}<0.05, \nonumber \\
\frac{|U_{e1}|^2}{|U_{e1}|^2-|U_{e2}|^2} &=& \cos^2{\Theta_{12}}, \;\;\;
\frac{|U_{e2}|^2}{|U_{e1}|^2-|U_{e2}|^2} = \sin^2{\Theta_{12}}.
\end{eqnarray}
That means that fitted parameters are the same as in 3-neutrino case
\begin{equation}
\sin^2{\Theta_{13}}=|U_{e3}|^2+|U_{e4}|^2 \equiv c_e<0.05.
\end{equation}
The MSW and VO solutions are described by $\sin^2{2\Theta_{12}}=\sin^2{2\Theta_{sun}}$ with the same values as 
in 3 neutrino scenario given in Table I.
For a scheme $B_4$ a change $U_{e3(4)} \leftrightarrow U_{e1(2)}$ must be done.
\subsection{Tritium beta decay}
Other constraints on neutrino masses and mixings come from the observation of the end of the Curie plot for the
tritium $\beta$ decay. Two collaborations from Mainz and Troitsk give similar results for the upper limit ($95 \%$
of c.l.) on the effective electron neutrino mass
\begin{equation}
\langle m_{\nu_e}\rangle _{\beta}=\left[ \sum\limits_{i=1}^n |U_{ei}|^2m_i^2 \right]^{1/2},
\end{equation}
\begin{eqnarray*}
\langle m_{\nu_e}\rangle _{\beta}&<&2.8 \;eV\;\;\; \mbox{\rm Mainz Collaboration } \cite{mainz}, \\
\langle m_{\nu_e}\rangle _{\beta}&<&2.5 \;eV\;\;\; \mbox{\rm Troitsk Collaboration } \cite{tro}.
\end{eqnarray*}
Both collaborations have ambitious plans to probe the mass region below 1 eV during the next five years \cite{Ref35}.

\subsection{Cosmological bounds}

There are also astrophysical and cosmological bounds on neutrino masses and mixings. All this information depends
on many other assumptions (as e.g. nonzero cosmological constant $\Lambda$) 
and is not as strict as laboratory data. We will take into account only one 
data which comes from the so called dark matter problem. If neutrinos compose all invisible matter in the
Universe then \cite{Ref40}
\begin{equation}
\sum m_{\nu} \leq 30 \;eV.
\end{equation}
If only 20 \% of all dark matter is formed by neutrinos (the so called Hot Dark Matter) then 
\begin{equation}
\sum m_{\nu} \simeq 6 \;eV.
\end{equation}
The best fit to many cosmological quantities  is obtained if around 70 \% of dark matter is given
by nonzero cosmological constant, 24 \% by Cold Dark Matter and 6 \% by Hot Dark Matter. In such a case 
 \begin{equation}
\sum m_{\nu} \simeq 2  \;eV.
\end{equation}

\section{Neutrinoless double beta decay and constraints on neutrino nature}

Neutrinoless double beta decay is sensitive to the first element of the neutrino mass matrix 

\begin{equation}
m_{\alpha \beta}=\sum\limits_{i=1}^n U_{\alpha i} U_{\beta i} m_i
\end{equation}
and luckily, is very well constrained.  This is not the case of other entries which are also measured
in various laboratory experiments, for instance

\begin{eqnarray} 
|m_{e \mu}| && \mbox{\rm in } Ti + \mu^- \to Ca + e^+ \nonumber \\
|m_{\mu \mu}| && \mbox{\rm in } K^+  \to \pi^- \mu^+  \mu^+ \\
|m_{e \tau }|,|m_{\mu \tau }|,|m_{\tau \tau }| && \mbox{\rm in HERA from}\;\; e^- p \to \nu_e l^- l'^-X.
\nonumber  
\end{eqnarray}
All these quantities have  quite large bounds, in the MeV-GeV range \cite{zub} e.g.
\begin{eqnarray} 
|m_{e \mu}| &<17 \;\;\; \mbox{\rm MeV}, \nonumber \\
|m_{\mu \mu}| &<500  \;\;\;  \mbox{\rm GeV}, \nonumber \\
|m_{e \tau}| &< 8.4  \;\;\; \mbox{\rm TeV}. 
\end{eqnarray}

The mixing matrix for Majorana neutrino has 3(6) phases for 3(4) neutrinos
so we have
\begin{eqnarray}
|\langle m_{\nu}\rangle |&=&\left| |U_{e1}|^2 m_1+ e^{2 i \phi_2} |U_{e2}|^2 m_2+e^{2 i \phi_3} |U_{e3}|^2 m_3 \right| , \\
|\langle m_{\nu}\rangle |&=&\left| |U_{e1}|^2 m_1+ e^{2 i \phi_2} |U_{e2}|^2 m_2+e^{2 i \phi_3} |U_{e3}|^2 m_3| 
+ e^{2 i \phi_4} |U_{e4}|^2 m_4 \right|, \nonumber \\ 
\end{eqnarray}
for $n=3(4)$ neutrinos, respectively.

We should stress that all our results are obtained in the approximation in which the lightest of neutrinos $(m_\nu)_{min}$
is heavier than
the difference of squares of neutrino masses responsible for solar neutrino oscillations $\left( (m_\nu)_{min} >> 
\delta m^2_{sun} \right)$.
\subsection{A schemes}

Let us first discuss the schemes $A_3$ and $A_4$. We have 
\begin{itemize}
\item for $A_3$
\begin{eqnarray}
m_1&=&(m_\nu)_{min}, \\
m_2&=& \sqrt{(m_\nu)_{min}^2+\delta m^2_{sun}} \simeq m_1 , \nonumber \\
m_3&=& \sqrt{(m_\nu)_{min}^2+\delta m^2_{atm}+\delta m^2_{sun}} \simeq  \sqrt{(m_\nu)_{min}^2+\delta m^2_{atm}}, \nonumber
\end{eqnarray}
\item for $A_4$
\begin{eqnarray}
m_1,m_2 &&  \mbox{\rm as for}\;\; A_3, \nonumber \\
m_3&=& \sqrt{(m_\nu)_{min}^2+\delta m^2_{LSND}+\delta m^2_{sun}} \simeq  \sqrt{(m_\nu)_{min}^2+\delta m^2_{LSND}}, \nonumber \\
m_4&=& \sqrt{m_3^2+\delta m^2_{atm}} \simeq m_3 .
\end{eqnarray}
\end{itemize}
Using the relation
\begin{equation}
min |z_1+z_2+z_3+z_4|= \left\{ 
\begin{array}{l}
|z_3+z_4|_{min}-|z_1+z_2|_{max}>0 , \cr
0, \cr
|z_1+z_2|_{min}-|z_3+z_4|_{max}>0 ,
\end{array}
\right.
\label{relmin}
\end{equation}
we get for both schemes
\begin{equation}
|\langle m_{\nu}\rangle |_{min}= \left\{ 
\begin{array}{ll}
s_{min}-\left( |U_{e1}|^2m_1+|U_{e2}|^2m_2 \right) & (m_\nu)_{min} \in (0,x_1^A), \cr
0 & (m_\nu)_{min} \in (x_1^A,x_2^A), \cr
\left| |U_{e1}|^2m_1-|U_{e2}|^2m_2| \right|-s_{max} & (m_\nu)_{min} > x_2^A,
\end{array}
\right.
\end{equation}
where 
\begin{eqnarray}
s_{\min }&=&s_{\max }=c_em_3,\;\;\;(A_3)\; \mbox{\rm scheme} \\
s_{\min }&=&\left| \left| U_{e3}\right| ^2m_3-\left| U_{e4}\right|
^2m_4\right| ,\;\;\;(A_4)\; \mbox{\rm scheme} \\
 s_{\max }&=&\left| U_{e3}\right| ^2m_3+\left| U_{e4}\right| ^2m_4,\;\;\;(A_4)\; \mbox{\rm scheme}.
\end{eqnarray}

$x_1^A$ and $x_2^A$ are the values of $(m_\nu)_{min}=m_1>0$ for which 
\begin{equation}
s_{\min }-\left( \left| U_{e1}\right| ^2m_1+\left| U_{e2}\right|
^2m_2\right) =0\mbox{ and }\left| \left| U_{e1}\right| ^2m_1-\left|
U_{e2}\right| ^2m_2\right| -s_{\max }=0
\end{equation}
respectively.

In both schemes there is (in agreement with Eq.~(\ref{st}) we take $\left|
U_{e1}\right| ^2>\left| U_{e2}\right| ^2$) 

\begin{equation}
\left| U_{e1}\right| ^2m_1+\left| U_{e2}\right| ^2m_2=(m_\nu)_{min} \left( 1-c_e\right) 
\end{equation}
and
\begin{equation}
\left|\left| U_{e1}\right| ^2m_1-\left| U_{e2}\right| ^2m_2\right|=(m_\nu)_{min} \left( 1-c_e\right) 
\sqrt{1-\sin ^22\theta _{sun}}.
\end{equation}
In the $A_4$ scheme we do not know $\left| U_{e3}\right| ^2$ and $\left|
U_{e4}\right| ^2$ separately and only approximate values for $s_{\max }$ can
be found 
\begin{equation}
s_{\max }=\left| U_{e3}\right| ^2m_3+\left| U_{e4}\right| ^2\sqrt{m_3+\delta
m_{atm}^2}\approx c_e\sqrt{(m_\nu)_{min}^2+\delta m_{LSND}^2}.
\end{equation}
The $s_{\min }$ is unknown so the region of $(m_\nu)_{min} \in \left( 0,x_1^A\right) $
cannot be checked precisely. We can find however that in both schemes 
$(\delta m^2=\delta m_{atm}^2$ or $\delta m_{LSND}^2)$ 
\begin{eqnarray}
&&\left| \left\langle m_\nu \right\rangle \right| _{\min }\\ &&\leq s_{\max
}-\left| U_{e1}\right| ^2m_1-\left| U_{e2}\right| ^2m_2=c_e\sqrt{(m_\nu)_{min}^2+\delta
m^2}-(m_\nu)_{min} \left( 1-c_e\right) \nonumber \\ &&<\left\{ 
\begin{array}{l}
c_e\sqrt{\delta m_{atm}^2}\approx 0.002\;\;\;eV\mbox{ for }A_3, \\ 
c_e\sqrt{\delta m_{LSND}^2}\approx 0.03\;\;\;eV\mbox{ for }A_4.
\end{array}
\right. \nonumber
\end{eqnarray}

The region $(m_\nu)_{min} >x_2^A$ is more interesting. In both schemes
this region occurs if
\begin{equation}
(m_\nu)_{min} \left( 1-c_e\right) \sqrt{1-\sin ^22\theta _{sun}}-c_e\sqrt{(m_\nu)_{min}^2+\delta m^2}%
\geq 0,
\end{equation}
from which the condition for $\sin ^22\theta _{sun}$ follows 
\begin{equation}
\label{condition}
\sin ^22\theta _{sun}\leq \frac{1-2c_e}{\left( 1-c_e\right) ^2}.
\end{equation}
For such values of mixing angle $\theta _{sun}$ we can find  $x_2^A$
 
\begin{equation}
x_2^A=\frac{\delta m^2c_e}{\sqrt{\left( 1-c_e\right) ^2\left( 1-\sin
^22\theta _{sun}\right) -c_e}}.
\end{equation}


\subsection{B schemes}

For $B_3$ and $B_4$ schemes the neutrino masses are connected with
the lightest neutrino mass as follows: 
\begin{itemize}
\item for $B_3$ 
\begin{equation}
\begin{array}{l}
m_1=(m_\nu)_{min}, \\ 
m_2=\sqrt{(m_\nu)_{min}^2+\delta m_{atm}^2}, \\ 
m_3=\sqrt{(m_\nu)_{min}^2+\delta m_{atm}^2+\delta m_{sun}^2}\approx m_3
\end{array}
\end{equation}
\item for $B_4$ 
\begin{equation}
\begin{array}{l}
m_1,m_2\mbox{ as in }B_3, \\ 
m_3=\sqrt{(m_\nu)_{min}^2+\delta m_{atm}^2+\delta m_{LSND}^2}, \\ 
m_4=\sqrt{m_3^2+\delta m_{sun}}\approx m_3.
\end{array}
\end{equation}
\end{itemize}
Using the relation Eq.~(\ref{relmin}) we obtain  
\begin{equation}
\left| \left\langle m_\nu \right\rangle \right| _{\min }=\left\{ 
\begin{array}{l}
w_{\min }-\left(\left| U_{e3}\right| ^2m_3+\left| U_{e4}\right| ^2m_4\right)>0,
\;\;\; (m_\nu)_{min}  \in \left( 0,x_1^B\right) \\ 
0, \;\;\; (m_\nu)_{min} \in \left( x_1^B,x_2^B\right) \\ 
\left|\left| U_{e3}\right| ^2m_3-\left| U_{e4}\right| ^2m_4\right|-w_{\max },
\;\;\; (m_\nu)_{min} >x_2^B.
\end{array}
\right.
\label{b}
\end{equation}
where  ($c_{e} =\left|
U_{e1}\right| ^2$ or $\left| U_{e1}\right| ^2+\left| U_{e2}\right| ^2)$%
\begin{eqnarray}
w_{\min }&=&w_{\max }=c_em_1 ,\;\;\;(B_3)\; \mbox{\rm scheme} \\ 
w_{\min }&=&\left| \left| U_{e1}\right| ^2m_1-\left| U_{e2}\right|
^2m_2\right| ,\;\;\;(B_4)\; \mbox{\rm scheme} \\
w_{\max }&=&\left| U_{e1}\right| ^2m_1+\left| U_{e2}\right|
^2m_2,,\;\;\;(B_4)\; \mbox{\rm scheme}.
\end{eqnarray}
$x_1^B$ and $x_1^B$ are solutions of the equations 
\begin{equation}
w_{\min }-\left(\left| U_{e3}\right| ^2m_3+\left| U_{e4}\right| ^2m_4\right)=0,\mbox{ and 
}\left|\left| U_{e3}\right| ^2m_3-\left| U_{e4}\right| ^2m_4\right|-w_{\max }=0,
\end{equation}
respectively.

Now there is (once more we assume $\left| U_{e3}\right| >\left|
U_{e4}\right| $) 

\begin{equation}
\left| U_{e3}\right| ^2m_3+\left| U_{e4}\right| ^2m_4=\left( 1-c_e\right) 
\sqrt{(m_\nu)_{min}^2+\delta m^2},\mbox{ and}
\end{equation}
\begin{equation}
\left| \left| U_{e3}\right| ^2m_3-\left| U_{e4}\right| ^2m_4 \right| =\left( 1-c_e\right) 
\sqrt{(m_\nu)_{min}^2+\delta m^2 \;\;\; }\sqrt{1-\sin ^22\theta _{sun}}
\end{equation}
where $\delta m^2=\delta m_{atm}^2$ (for $B_3)$ and $\delta m^2=\delta
m_{atm}^2+\delta m_{LSND}^2$ for $B_4$.
As $0\leq c_e\leq 0.05$ 
\begin{equation}
w_{\min }-\left( \left| U_{e3}\right| ^2m_3+\left| U_{e4}\right|
^2m_4\right) <c_e (m_\nu)_{min}- \left( 1-c_e\right) \sqrt{(m_\nu)_{min}^2+\delta m_{atm}^2}<0,
\end{equation}
and  the first two regions in Eq.~(\ref{b}) are not
present.
In the $B_4$ scheme, as in the $A_4$, we do not know $\left| U_{e1}\right| ^2$ and $%
\left| U_{e2}\right| ^2$ separately.

For $w_{\max }$ only the bound can be found 
\begin{equation}
c_e (m_\nu)_{min} <w_{\max }<c_e\sqrt{(m_\nu)_{min}^2+\delta m_{atm}^2}.
\end{equation}
In this case the $\left| \left\langle m_\nu \right\rangle \right| $
satisfies 
\begin{equation}
\left| \langle m_\nu \rangle \right| _{\min }\geq \left( 1-c_e\right) \sqrt{(m_\nu)_{min}^2+\delta m^2%
}\sqrt{1-\sin ^22\theta _{sun}}-c_em_{min}\equiv f\left[(m_\nu)_{min} \right]
\label{fx}
\end{equation}

where $m_{min}=(m_\nu)_{min}$ for $B_3$ and $m_{min}=\sqrt{(m_\nu)_{min}^2+\delta m_{atm}^2}$ for $%
B_4.$

If condition Eq.~(\ref{condition}) is satisfied $f\left[ (m_\nu)_{min} \right] $ is an increasing
function of $(m_\nu)_{min}$, if not, $f\left[ (m_\nu)_{min}\right] $ decreases from 
\begin{equation}
f\left[ 0\right] =\left( 1-c_e\right) \sqrt{\delta m^2}\sqrt{1-\sin
^22\theta _{sun}}-c_em_{min}\left[ (m_\nu)_{min}=0\right] \mbox{ for }(m_\nu)_{min}=0
\end{equation}
to
\begin{equation}
f\left[ (m_\nu)_{min}\right] =0\mbox{ for }(m_\nu)_{min}=\left[ \frac{\delta m^2\left( 1-\sin
^22\theta _{sun}\right) -\frac{c_e^2}{\left( 1-c_e\right) ^2}m_{min}\left[
(m_\nu)_{min}=0\right] }{\sin ^22\theta _{sun}-\frac{\left( 1-2c_e\right) }{\left(
1-c_e\right) ^2}}\right] ^{1/2}.
\end{equation}



All the above analytical considerations lead us to the following conclusions (for plots see \cite{Ref10}).
\begin{enumerate}
\item[A)] Present bound $\left| \left\langle m_\nu \right\rangle \right| <0.2$ $eV.$ (see 
Fig.~\ref{present_and_genius1})
\begin{figure}[h]
\epsfig{figure=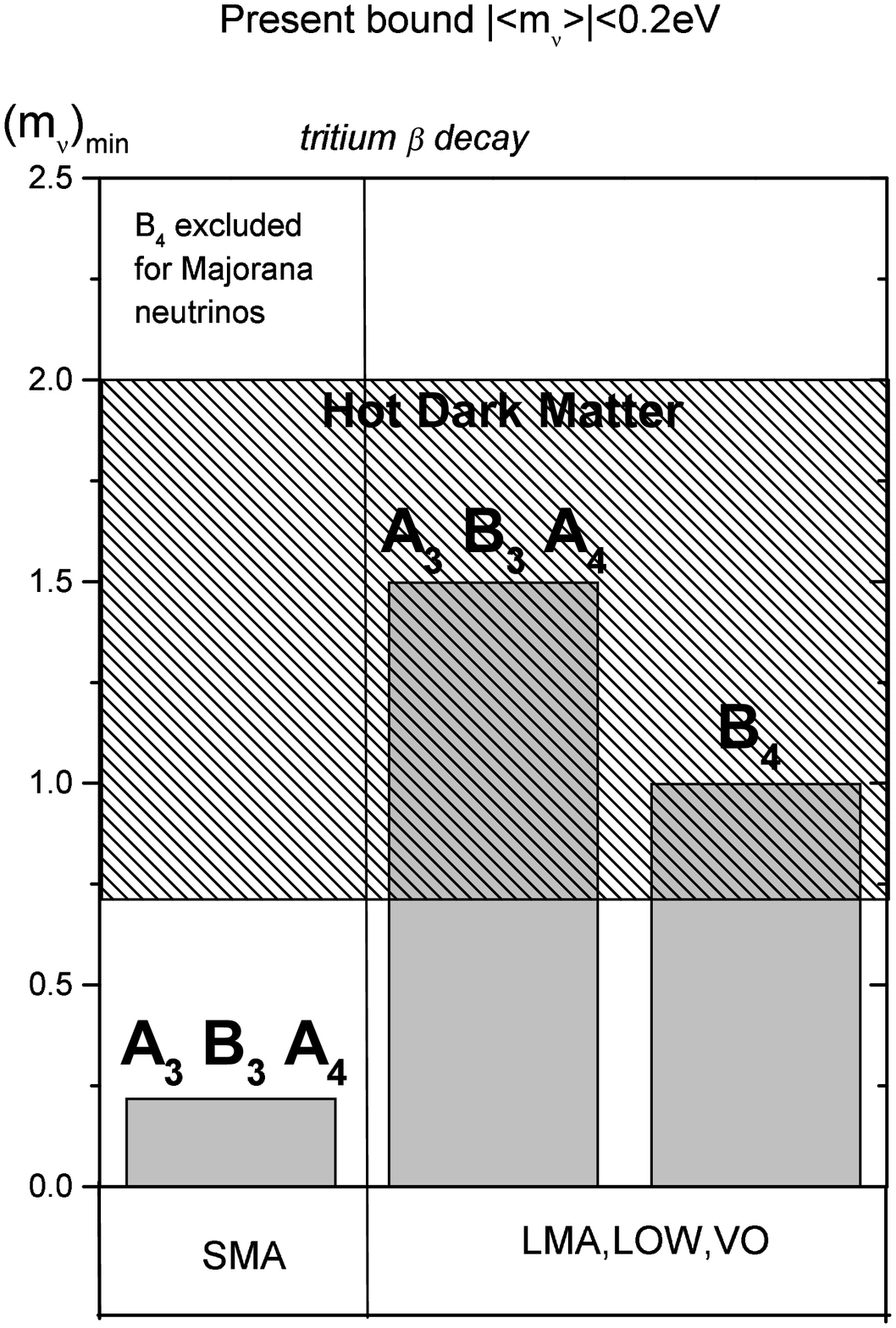,width=7cm}
\epsfig{figure=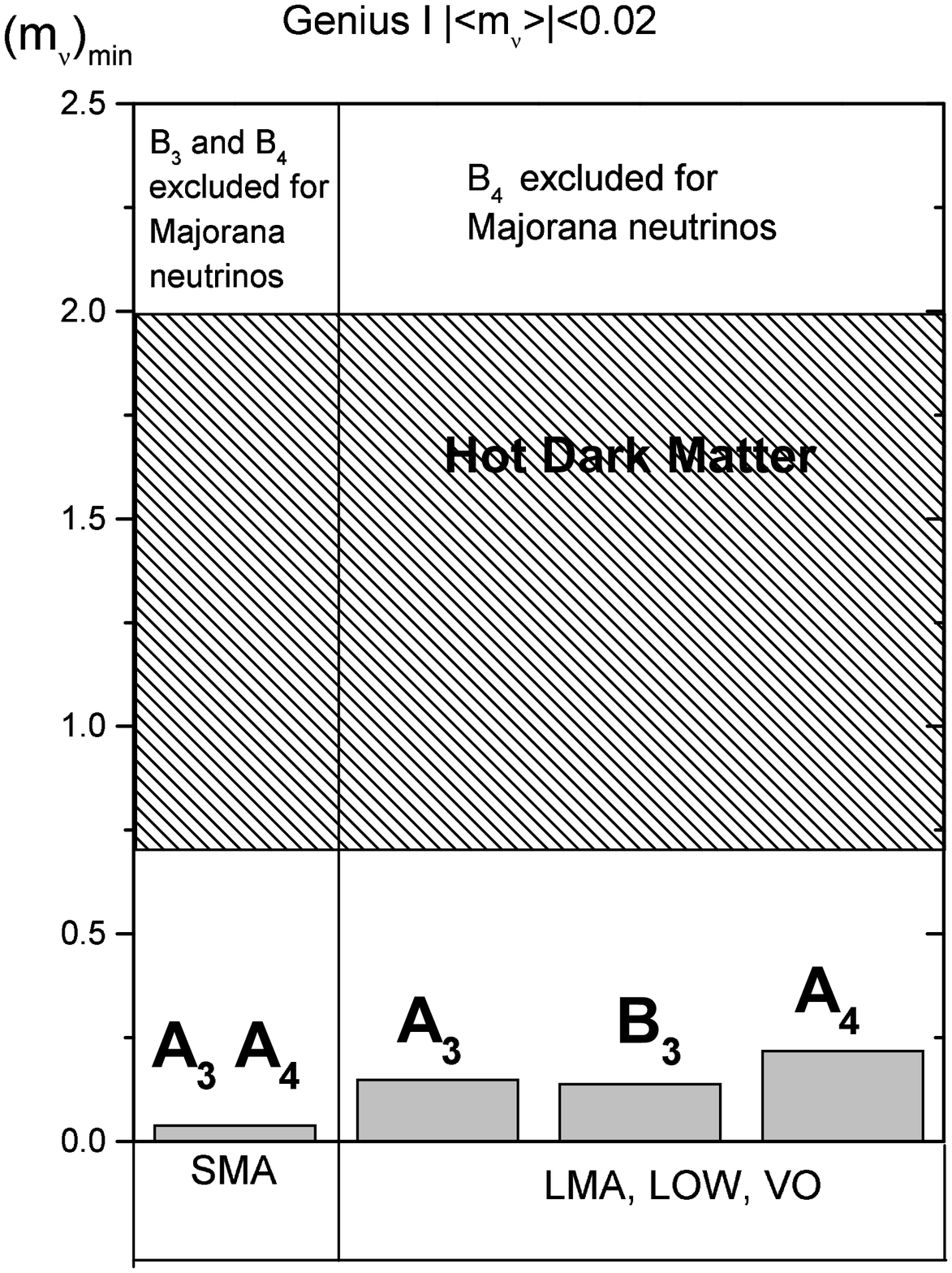,width=7cm}
\caption{Upper limits on the mass of the lightest of Majorana neutrinos derived from present (left) and GENIUS I (right)
$(\beta\beta)_{0\nu}$ experimental bounds for different
neutrino mass schemes and various solar neutrino oscillation
solutions. The gray shaded area shows the allowed mass region  for
this neutrino. The HDM area applies in the three neutrino case only. We
can see that Genius I with HDM solve the problem of neutrinos' nature in
this case.
\label{present_and_genius1}}
\end{figure}
\begin{itemize}
\item \underline{If SMA MSW solution is the proper mechanism then:}
\begin{itemize}
\item $B_4$ scheme is excluded for Majorana neutrinos,
\item In schemes $A_3,A_4$ and $B_3$ Majorana neutrinos are accepted if $\left(
m_\nu \right) _{\min }<0.22$ $eV.$ Above this mass all three schemes are
open only for Dirac neutrinos.
\end{itemize}
\item \underline{For LMA and LOW MSW solutions:}
\begin{itemize}
\item the $A_3,A_4$ and $B_3$ schemes accept Majorana neutrinos only for $\left(
m_\nu \right) _{\min }<1.5$ $eV$, an analogous limit in the $B_4$ scheme is $%
\left( m_\nu \right) _{\min }<1.1$ $eV.$
\end{itemize}
\end{itemize}
\item[B)] If GENIUS I gives only a bound $\left| \left\langle m_\nu
\right\rangle \right| <0.02$ $eV:$ (see Fig.~\ref{present_and_genius1})
\begin{itemize}
\item \underline{For SMA MSW solution:}
\begin{itemize}
\item scheme $B_3$ is excluded for Majorana neutrinos,
\item in schemes $A_3$ and $A_4$ Majorana neutrinos are accepted only for small
masses $\left( m_\nu \right) _{\min }<0.04$ $eV.$
\end{itemize}
\item \underline{For LMA and LOW solutions:}
\begin{itemize}
\item the $B_4$ scheme is excluded for Majorana neutrinos,
\item Majorana neutrinos can exist for $\left( m_\nu \right) _{\min }<0.16$ $eV$ 
$\left( A_3\right) ,$ $\left( m_\nu \right) _{\min } <0.14$ $eV$ $\left( B_3\right) $ and $\left( m_\nu \right) _{\min }
<0.22eV\left( A_4\right) .$
\end{itemize}
\end{itemize}
\item[C)] If finally GENIUS II does not find the $\left( \beta \beta
\right) _{0\nu}$ decay (see Fig.~\ref{genius2}):
\begin{center}
\begin{figure}[h]
\epsfig{figure=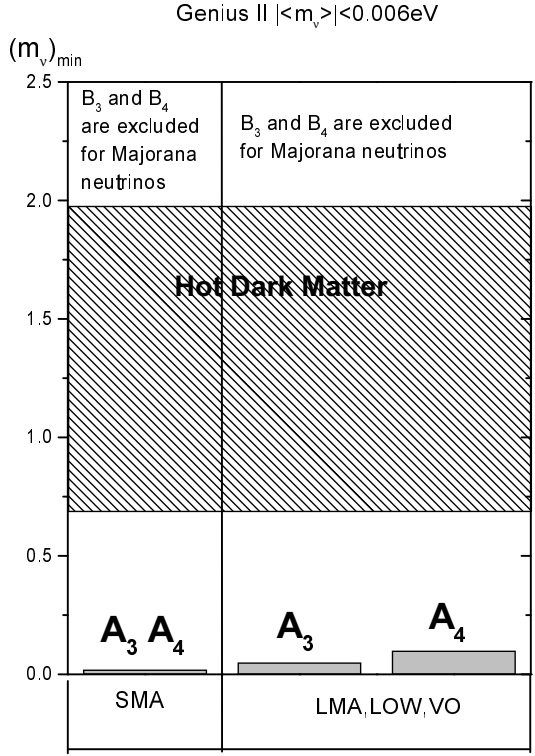,width=8.5cm}
\caption{Even more restricted area allowed for the mass of the lightest Majorana neutrino in the case of GENIUS II.
\label{genius2}}
\end{figure}
\end{center}
\begin{itemize}
\item \underline{For SMA MSW solution:}
\begin{itemize}
\item Majorana neutrinos with $\left( m_\nu \right) _{\min}\leq 0.02eV$ can exist
only in the $A_3$ and $A_4$ schemes
\end{itemize}
\item \underline{For LMA MSW and LOW solutions}
\begin{itemize}
\item the $B_3$ scheme is excluded for Majorana neutrinos,
\item Majorana neutrinos can exist only in $A_3$ and $A_4$ schemes with $\left(
m_\nu \right) _{\min }<0.05$ eV and $\left( m_\nu \right) _{\min}<0.12$ eV,
respectively.
\end{itemize}
\end{itemize}
\end{enumerate}

There are additional restrictions with assumption that neutrinos contribute
to the dark matter content. Three  neutrinos with almost degenerate
masses $m_\nu \sim 0.7$ $eV$ $\left( 2eV\right) $ must exist if $\sum m_\nu
\approx 2$ $eV$ $\left( 6\;\;\; eV\right) .$ This means that already the present $\left( \beta \beta \right) _{0\nu }$ bound
closes all schemes for three Majorana neutrinos if the SMA solution is the
proper one. The GENIUS\ I bound will close schemes for three Majorana
neutrinos.
For schemes $A_4$ and $B_4$
with a sterile neutrino $\left( m_\nu \right) _{\min }$ must be very small if  $\sum m_\nu \simeq 2 \;eV$ and
 $\left( m_\nu \right) _{\min }\approx 1.0$ if 
$\sum m_\nu \approx 6$ $eV$.
 Then the only scheme  with one sterile neutrino is
accepted if the sum of all Majorana neutrinos is approximately 2 eV.
If $\sum m_\nu \sim 6$ $eV$ and GENIUS I will give negative results only 3
or 4 Dirac neutrinos can constitute the HDM. In such case there is a problem
how to explain the number of neutrino degrees of freedom from the abundance
of the $^4$He and D/N. The present highest bound is $N_\nu <5.3$ \cite{Ref43}.

\section{Conclusions.}

We have entered an exciting era in neutrino physics. Mixing in the lepton
sector seems to be established. An obvious consequence of this fact is the
nonconservation of lepton family number $L_\alpha .$ Breaking of the $%
L_\alpha $ is very weak. It is seen only in neutrino oscillation and in
no other terrestrial laboratory experiments. The problem of the conservation or violation of the total 
lepton number L, 
which is connected with the Dirac or Majorana neutrino
nature, is not solved up to now. Approximate conservation of $L_\alpha $ and 
$L$ follows from i) smallness of neutrino masses, ii)ultrarelativistic
character of produced neutrinos, iii) unitarity (exact or approximate) of
the mixing matrix, iv) left-handed nature of the neutrino interaction.

For Majorana neutrinos this approximate $L_\alpha $ and $L$ conservation can
be proved even though lepton numbers are not defined for neutral particles.

The Majorana neutrino mass matrix elements $m_{\alpha \beta }$ ($\alpha
,\beta =e,\mu ,\tau $) are bounded by various experiments. Such bounds are
usually in the MeV-GeV range. Only one element $m_{ee}$ is limited
with good enough precision to play a role in the reconstruction of
the mixing in the lepton sector. The $m_{ee}$ element is measured in double $\beta $
decay of various nuclei. Up to now this decay has not been observed. The contrary would establish the 
Majorana nature of neutrinos. However the combination of
various informations about masses and mixing matrix elements from i)
oscillation experiments, (ii) tritium $\beta $ decay and (iii) cosmology
together with $\left( \beta \beta \right) _{0\nu }$ is able to discriminate
between the accepted neutrino mass spectra allowed for Majorana or only
for Dirac neutrinos. The data are not precise enough to make  conclusive
statements. The bound on $\left\langle m_\nu \right\rangle $depends strongly
on the determination of nuclear matrix elements. Our estimation was made
with 95\% CL. At $3\sigma $ which corresponds to 99\% CL, a value of one for $\sin
^22\theta _{sun}$ is accepted and we cannot make any discrimination between
the two natures.

Our estimation is interesting also for those who strongly believe that
neutrinos are Majorana particles. We found the corner of the mass schemes
where such neutral particles are still allowed. With the present experimental
precision the room for the Majorana neutrino is bounded but still large. If
next $\left( \beta \beta \right) _{0\nu }$ experiments give negative results
the Majorana neutrino corner will become smaller and smaller. More precise
informations about i) existence of sterile neutrino, ii) which solution of
the solar neutrino anomaly is accepted and iii) knowledge of oscillation
parameters with smaller error are urgently needed. We hope that future
(already working and planned) experiments will provide us with these
informations and together with the neutrino mass scheme, the  neutrino
character will be established.

{}
\end{document}